\documentclass[%
 aip,
 amsmath,amssymb,
 reprint,%
]{revtex4-1}

\usepackage{amssymb}
\usepackage{color,graphicx}
\usepackage{lmodern}
\usepackage{amsmath}
\usepackage{amsbsy}
\usepackage{float}
\usepackage{bbm}
\usepackage{bm}
\usepackage{epsfig}
\usepackage{braket}
\usepackage{graphics}
\usepackage{subfigure}
\usepackage{appendix}
\usepackage{lipsum}
\usepackage{relsize}

\usepackage[none]{hyphenat}

\newcommand\EXP{\operatorname{exp}}

\newcommand\avg{\operatorname{avg}}

\newcommand\thh{\operatorname{th}}
\newcommand\Int{\operatorname{int}}
\newcommand\Spon{\operatorname{spon}}
\newcommand\PD{\operatorname{PD}}
\newcommand\CS{\operatorname{CS}}
\newcommand\QS{\operatorname{QS}}
\newcommand\qq{\bf q}
\newcommand\dd{\bf{d}}
\newcommand\rr{\bf{r}}
\newcommand\kk{\bf{k}}
\newcommand\eeff{\operatorname{eff}}
\newcommand\ccvv{\it{ul}}
\newcommand\Photon{\operatorname{photon}}
\newcommand\ccm{\operatorname{cm}}
\newcommand\e{\operatorname{e}}
\newcommand\ee{\bf{e}}
\newcommand\ccos{\operatorname{cos}}
\newcommand\Prr{\operatorname{Pr}}
\newcommand\ex{\operatorname{ex}}
\newcommand{\mycomment}[1]{}

\begin{document}
\title{Quantum Random Number Generator Based on LED}

\author{Mohammadreza Moeini}
\affiliation{Iranian Quantum Technologies Research Center (IQTEC), Tehran, Iran}

\author{Mohsen Akbari}
\email[Corresponding Author:~]{mohsen.akbari@khu.ac.ir}

\affiliation{Quantum Optics Lab, Department of Physics, Kharazmi University, Tehran, Iran}

\author{Mohammad Mirsadeghi}
\affiliation{Department of Electrical Engineering, K. N. Toosi University of Technology, Tehran, Iran}

\author{Hamid Reza Naeij}
\affiliation{Department of Chemistry, Sharif University of Technology, Tehran, Iran}

\author{Nima Haghkish}
\affiliation{Iranian Quantum Technologies Research Center (IQTEC), Tehran, Iran}

\author{Ali Hayeri}
\affiliation{Iranian Quantum Technologies Research Center (IQTEC), Tehran, Iran}

\author{Mehrdad Malekian}
\affiliation{Department of Electrical Engineering, Sharif University of Technology, Tehran, Iran}

\begin{abstract}

Quantum Random Number Generators (QRNGs) produce random numbers based on the intrinsic probabilistic nature of quantum mechanics, making them True Random Number Generators (TRNGs).  In this paper, we design and fabricate an embedded QRNG that produces random numbers based on fluctuations of spontaneous emission and absorption in a Light-Emitting Diode (LED). To achieve a robust and reliable QRNG, we compare some usual post-processing methods and select the finite impulse response (FIR) method for a real-time device. This device could pass NIST tests, the generation rate is 1 Mbit/s and the randomness of the output data is invariant in time.

\end{abstract}

\maketitle

This century can be considered the beginning of the rapid development and dissemination of quantum information technology in almost all scientific and utility fields. Meanwhile, random numbers play a significant role in many aspects of information technology ~\cite{Meteopolis, Bennett_Brassard, Schneier}. Applications of random numbers including symmetric key cryptography~\cite{Metropolis_2}, Monte Carlo simulation~\cite{Dynes}, transaction protection~\cite{Zhang}, and key distribution systems~\cite{Symul} will be more important in the era of quantum technology. 

Traditionally, pseudo-random number generators (pseudo-RNGs) were based on deterministic algorithms and could not generate truly random numbers. On the other hand, quantum random number generators (QRNGs) can generate truly random numbers from the essentially probabilistic nature of quantum processes and can also provide higher bit rates instead of physical random number generators~\cite{Schmidt}. 
 
To date, various practical protocols for QRNGs have been proposed, such as QRNG-based photon counting~\cite{Nie, Jennewein, Dynes}, quantum entanglement~\cite{JJacak}, Raman scattering~\cite{Bustard_Raman}, vacuum fluctuations~\cite{Shen, Gabriel}, amplified spontaneous emission~\cite{Li}, radioactive decay~\cite{Alkassar}, and laser phase noise~\cite{Nie, Qi}. References~\cite{Mannalath, Ma, Jacak} are available to the reader for further information.

In this work, we experimentally demonstrate a simple, inexpensive, and real-time QRNG based on the spontaneous emission and absorption of an LED and easily accessible at high speed.
  
First, we discuss spontaneous emission in an LED-QRNG device, which is widely recognized as the primary source of randomness. Following this, we provide a theoretical formalism that explains absorption as a secondary mechanism modifying output light, and demonstrate that absorption and spontaneous emission both contribute to randomness. We will then demonstrate the physical setup and illustrate an approach to maximize the quantum part of the signal. Finally, the post-processing procedures and results are discussed.

\begin{figure*}
\centering
\begin{subfigure}[]{
\label{fig_Fermi_recombinations}
\centering
\includegraphics[scale=0.95]{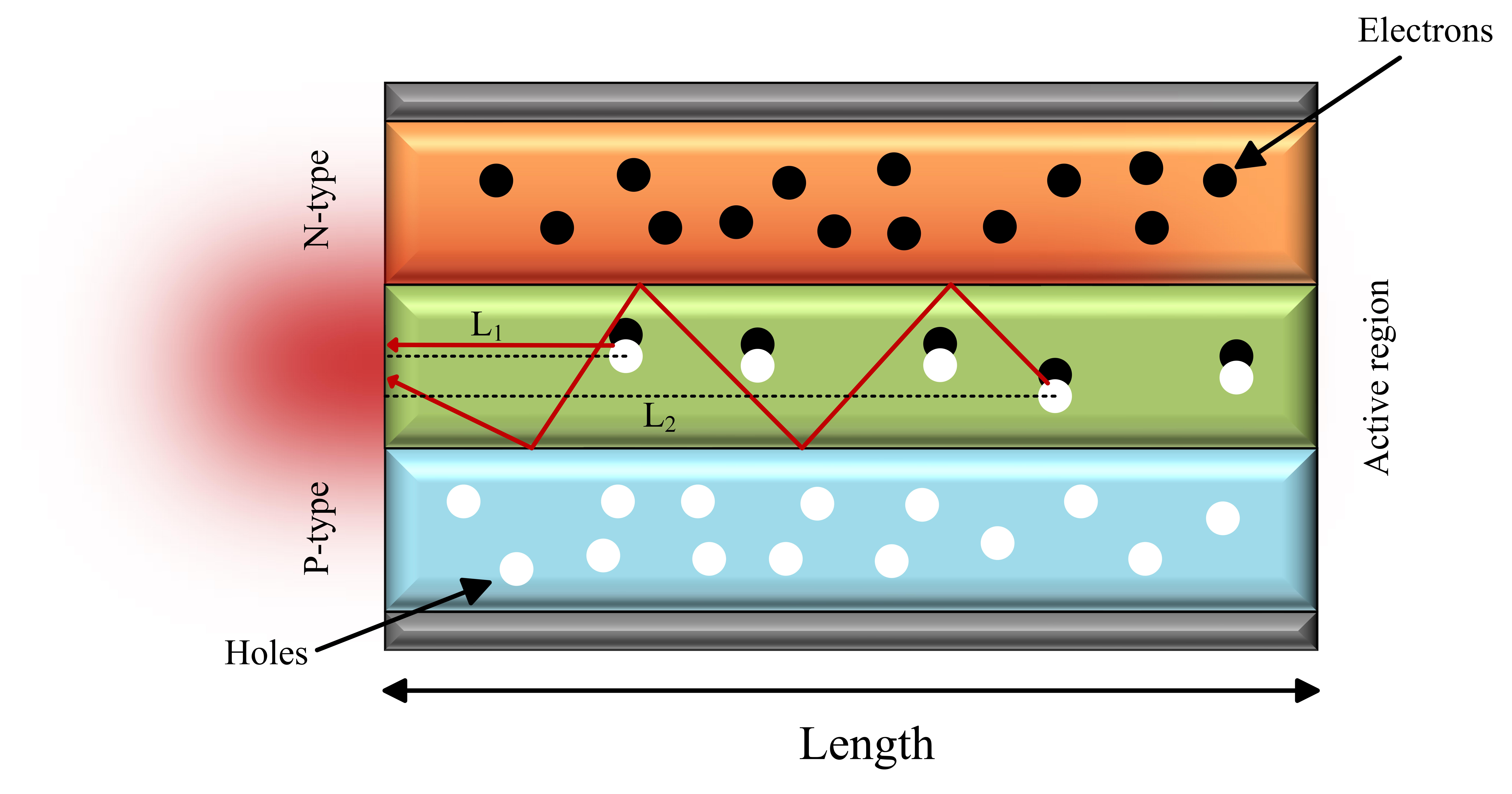}}
\end{subfigure}
\begin{subfigure}[]{
\label{fig_QRNG}
\includegraphics[scale=0.11]{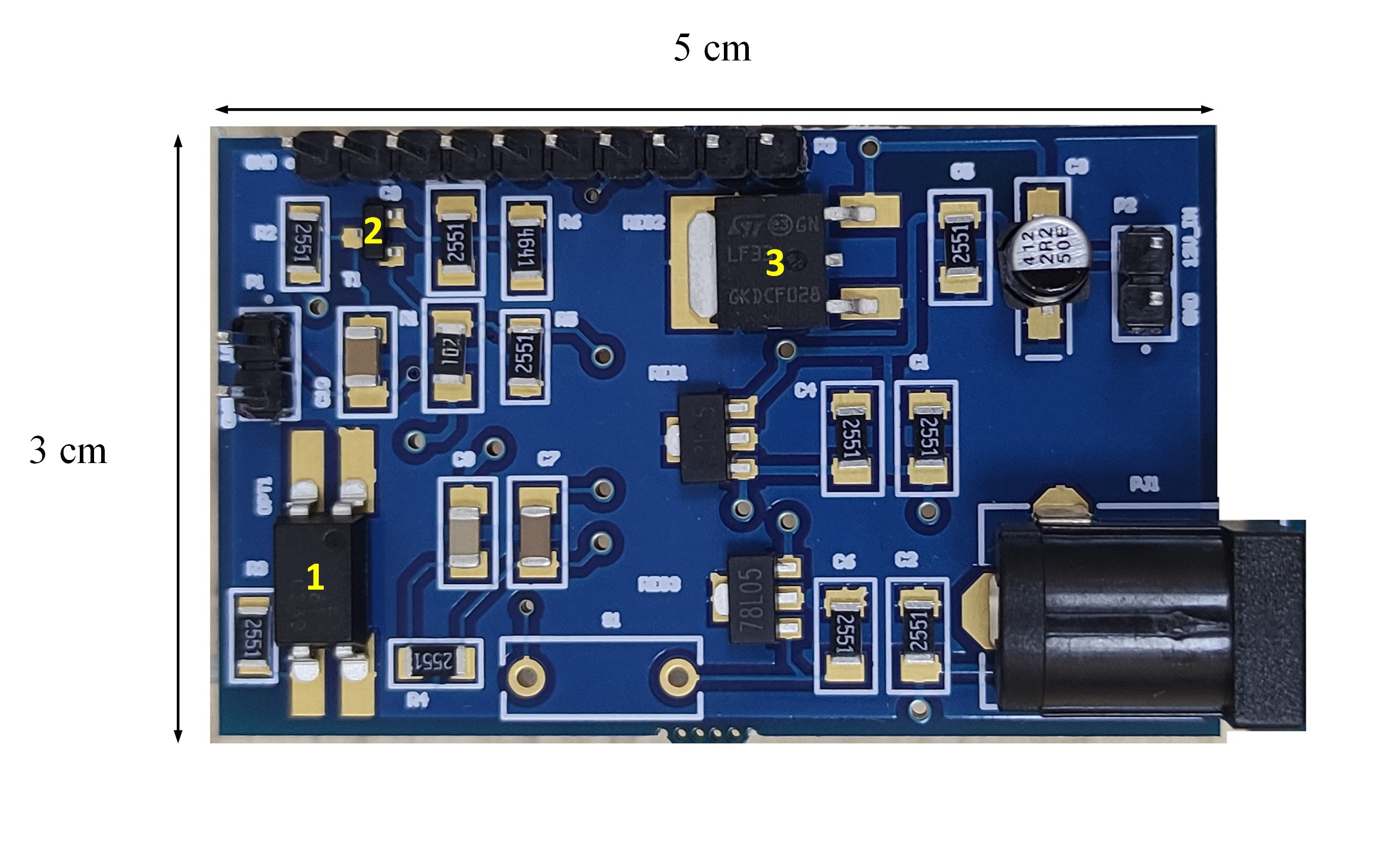}}
\end{subfigure}
\begin{subfigure}[]{
\label{fig_Schematic}
\centering
\includegraphics[scale=0.98]{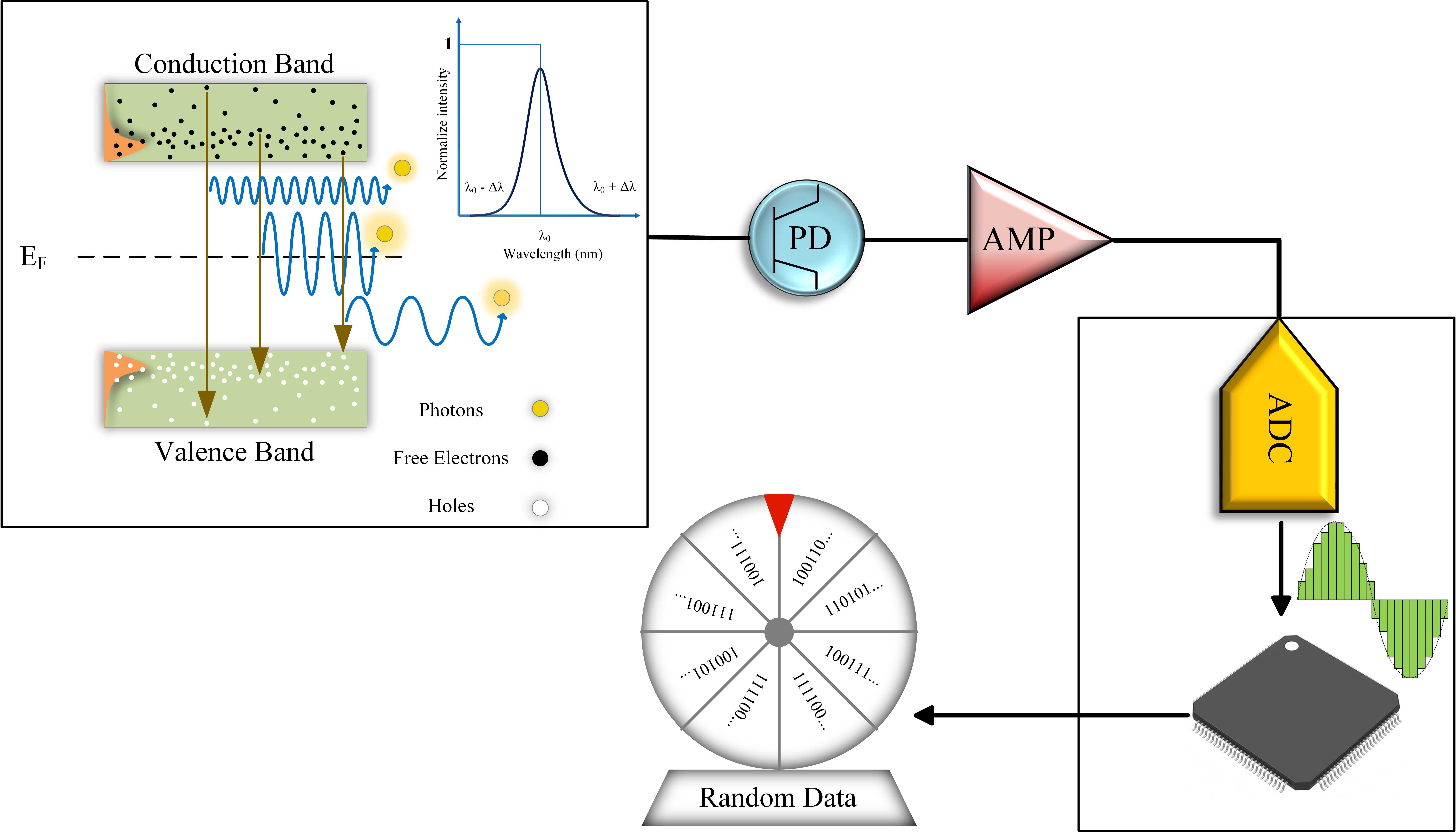}}
\end{subfigure} 

\caption{(a) The process of light emission from an LED involves the application of a suitable voltage, which causes electrons and holes to recombine randomly in the active region. It is worth noting that the length of output for each photon is different. (b) The QRNG’s PCB. The LED and detector are integrated into an opto-isolator (component 1). Component 2 is the BJT amplifier and component 3 is the regulator. (c) Diagram of QRNG. The LED's energy diagram illustrates the density of state for electrons and holes, with a varying absorption coefficient for each photon due to different frequencies. After emission, the photon is absorbed by the photodiode, and the output electrical signal is amplified, then digitized by an ADC and processed through a microprocessor.}
\end{figure*}

In an isotropic optical medium, the total spontaneous recombination rate $R_{\Spon} $ can be defined as~\cite{Lasher}
\begin{align}
\label{Eq_spontaneous_emission}
R_{\Spon} = \int{{\rr}_{\Spon}(\mathcal{E})d\mathcal{E}}
\end{align}
where ${\rr}_{\Spon}(\mathcal{E})$ denotes the spontaneous downward transitions in an electronic system and written as
\begin{align}
\label{Eq_spontaneous_emission}
{\rr}_{\Spon}(\mathcal{E})d\mathcal{E} = \mathlarger{\Sigma}{\Big(4n_{b}e^{2}\mathcal{E}/m^{2}\hbar^{2}c^{3}\Big)} |M|^{2} f_{u} (1-f_{l}) d\mathcal{E}
\end{align}
in which $f_{u}$ and $f_{l}$ are the probabilities that the upper and lower states involved in the transition are occupied, $n_{b}$, $m$, $e$, and $c$ are the refractive index, effective mass, electron charge, and the speed of light, respectively and the sum is taken over all pairs of states per unit volume whose energy difference is between $\mathcal{E}$ and $d\mathcal{E}$. Moreover, the matrix element is averaged over all polarizations of the incident light, so, we can write
\begin{align}
\label{matrix_element}
|M|^{2} = \frac{1}{3}{|M_{x}^{2} + M_{y}^{2} + M_{z}^{2}|},\nonumber\\
M_{j} = -i\hbar \big(\psi_{u} \mid \EXP(i\kk \cdot \rr)(\frac{\partial}{\partial {\textnormal j}})\mid \psi_{\it{l}}\big)
\end{align}
where $j=x,y,z$. Furthermore, $\psi_{u}$ and $\psi_{l}$ are the wave functions of the upper and lower states, and $\kk$ is the propagation vector of the radiation. After some derivation, we get the formula for the spontaneous recombination rate as
\begin{align}
\label{R_SPON}
R_{\Spon} = Bnp
\end{align}
where $n$ and $p$ are the concentrations of electrons and holes and $B$ is given by~\cite{Lasher}
\begin{align}
\label{B_matrix}
B = (4n_{b}e^2\mathcal{E}/m^2\hbar^{2}c^{3})\langle |M|^2\rangle_{\avg} V
\end{align}
where $\langle |M|^2\rangle_{\avg}$ is the average of the squared matrix element over spins in the upper and lower bands, and $V$ is the volume of the crystal. Further, we will see that $R_{\Spon}$ is the source of randomness due to the emission.

As discussed earlier, spontaneous emission is the primary mechanism for emitting light from an LED, and it is a quantumly random process~\cite{Henry, Xu}. Aside from that, we propose that absorption is a quantum entropy source in LEDs alongside spontaneous emission.

Beer-Lambert's law can be used as a starting point to show that absorption fluctuation affects light intensity as well. As shown in FIG.~\ref{fig_Fermi_recombinations}, the electron-hole recombination process generates the light with intensity $I_{0}$. Light may be absorbed when it passes through the active region that is expressed as
\begin{align}
\label{Eq_Beer_Lambert}
I = I_{0}e^{-\alpha L}
\end{align}
where $\alpha$ represents the absorption coefficient of the object and $L$ denotes the distance traveled by the light. Absorption occurs when an electron is excited by a photon. Despite the fact that Beer-Lambert law is applicable in the macroscopic region since light is treated as classical electromagnetic waves, it can be extended to the microscopic domain if one considers photon number decay due to the absorption of photons during the process~\cite{Quimby}.
Thus, we use time-dependent perturbation in order to determine the absorption coefficient~\cite{Nasser, Casey}.

The interaction Hamiltonian for a generated photon into the active region is~\cite{Nasser}:
\begin{align}
\label{Eq_Hint_photon}
H_{\Int}^{\Photon} = -E_{0}{\ccos}({\qq} {\rr} - \omega t) {\ee}_{q} {\dd} 
\end{align}
where $E_{0}$, $\qq$, $\omega$, and ${\ee}_{q}$ are the electric field amplitude, wavenumber, angular frequency, and polarization unit vector of the generated photon, respectively, and $\dd = -\e\rr$ denotes the light-induced electric field dipole moment. Considering the dipole approximation, we obtain:
\begin{align}
\label{Eq_dipole_appx}
\langle \psi_{u} | H_{\Int}^{\Photon} | \psi_{l}\rangle = -\frac{E_{0}}{2} (e^{-i \omega t} + e^{i \omega t}) {\dd}_{\ccvv}
\end{align}
which ${\dd}_{\ccvv} = \langle \psi_{u} |{\ee}_{q} {\dd}| \psi_{l} \rangle$ is the dipole element for transition between $\psi_{u}$ and $\psi_{l}$ states. Upon solving the Schrödinger equation with a perturbation harmonic in time for $H_{\Int}^{\Photon}$, assuming an electron transitions from initial state $l$ to final state $u$, the probability is~\cite{Nasser, Casey}:
\begin{align}
\label{Eq_FermiGoldenRule}
w_{\ccvv}=\Big \lvert \frac{1}{i\hbar}\int_{0}^{t} \langle\psi_{u}|H_{\Int}^{\Photon}|\psi_{l}\rangle e^{i\omega_{\ccvv}t'}dt' \Big\rvert ^2
\end{align}
where $\frac{2\pi}{\hbar}|\langle\psi_{u}|H_{\Int}^{\Photon}|\psi_{l}\rangle|^{2}$ refers to the transition rate probability, also known as Fermi's golden rule~\cite{Casey}, which is the basis for the randomness in absorption.
Additionally, the absorption coefficient can be defined as~\cite{Chaung}: 
\begin{align}
\label{Eq_abs_coeff}
\alpha = \frac{w_{\ccvv}\hbar\omega}{tS}
\end{align}
where $S$ represents the Poynting vector, and $w_{\ccvv}/t$ is the transition probability per unit of time or the transition rate probability. Regarding Eqs.~(\ref{Eq_Hint_photon})-(\ref{Eq_abs_coeff}), one can derive $\alpha$ as~\cite{Nasser}:

\begin{align}
\label{Eq_abs_coeff2}
\alpha = \frac{\omega}{\pi n_{b} c} | {\dd}_{\ccvv}|^2  \int_{0}^{\infty} 4\pi k^2 \delta(\hbar \omega_{\ccvv} - \hbar \omega) ~dk 
\end{align}
Eq.~(\ref{Eq_abs_coeff2}) gives the absorption coefficient in terms of the density of states and the dipole element.

For any LED which $\alpha L \ll 1$, we can approximate Eq. (\ref{Eq_Beer_Lambert}) as~\cite{Chaung}:

\begin{align}
\label{Eq_Beer_Lambert_appx}
I \approx I_{0}(1 - \alpha L)
\end{align}
In other words, the light intensity fluctuation includes fluctuations of $I_{0}$, $\alpha$, and $L$. Considering  Eqs.~(\ref{Eq_FermiGoldenRule}),(\ref{Eq_abs_coeff}), the $\alpha$ depends on the transition rate probability and the energy of the generated photons. These terms are probabilistic and cause fluctuations in the light intensity. For example, wavelength broadening has been shown in FIG. \ref{fig_Schematic}, which can directly cause a variation in the light intensity. For simplicity, we can average the $L$ along the cavity and take it as a constant. Considering this, we define effective length $L_{\eeff}$
\begin{align}
\label{Eq_L_avg}
\int_{0}^{L_{0}}{I_{0}e^{-\alpha x}}dx = I_{0} L_{\eeff}
\end{align}
where $x$ represents the position of the generated photons and $L_{0}$ represents the active region length. After some calculation, we can reach
\begin{align}
\label{Eq_L_avg_2}
L_{\eeff} = \frac{1}{\alpha} (1-e^{-\alpha L_{0}}) \approx L_{0}
\end{align}
Consequently, we no longer have to consider $L$ as a random variable. 

\begin{figure*}
\centering
\begin{subfigure}[]{
	\centering
	\label{fig_time_wave}
	\includegraphics[scale = 0.35]{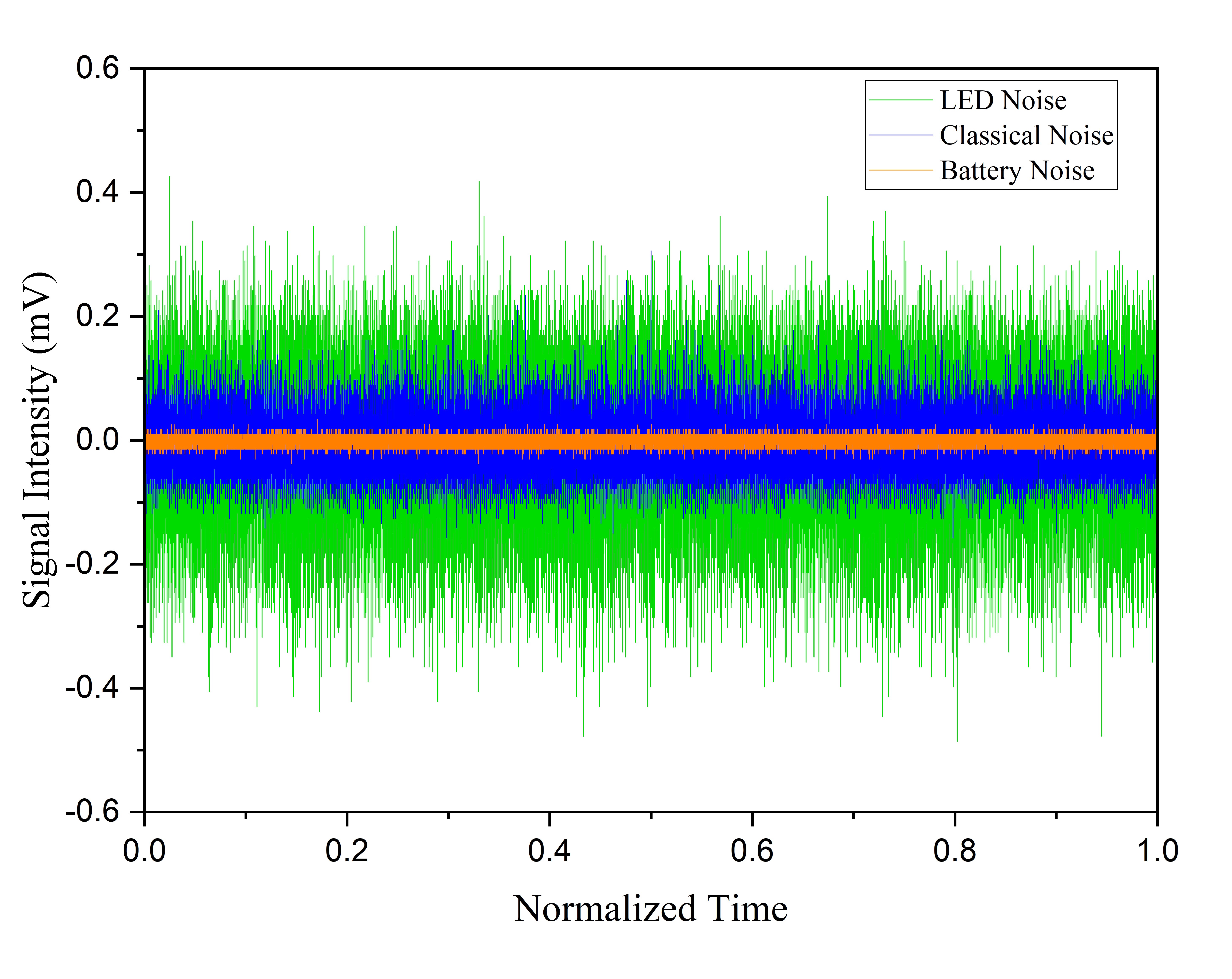}}
	\end{subfigure}
	\begin{subfigure}[]{
		\centering
		\label{fig_hist_wave}
		\includegraphics[scale = 0.35]{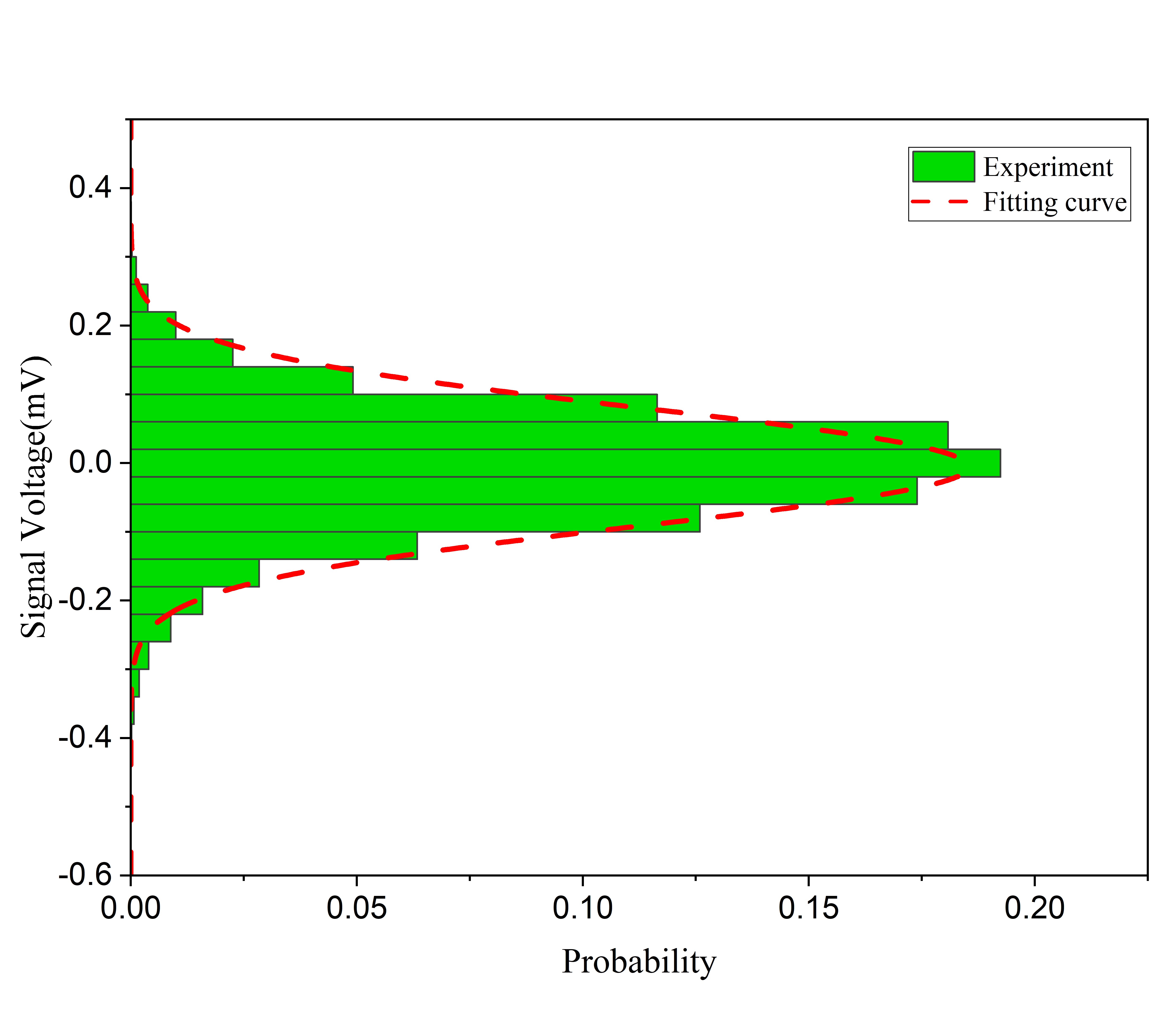}}
		\end{subfigure}
	\begin{subfigure}[]{
		\label{FFT_Power}
		\includegraphics[scale = 0.35]{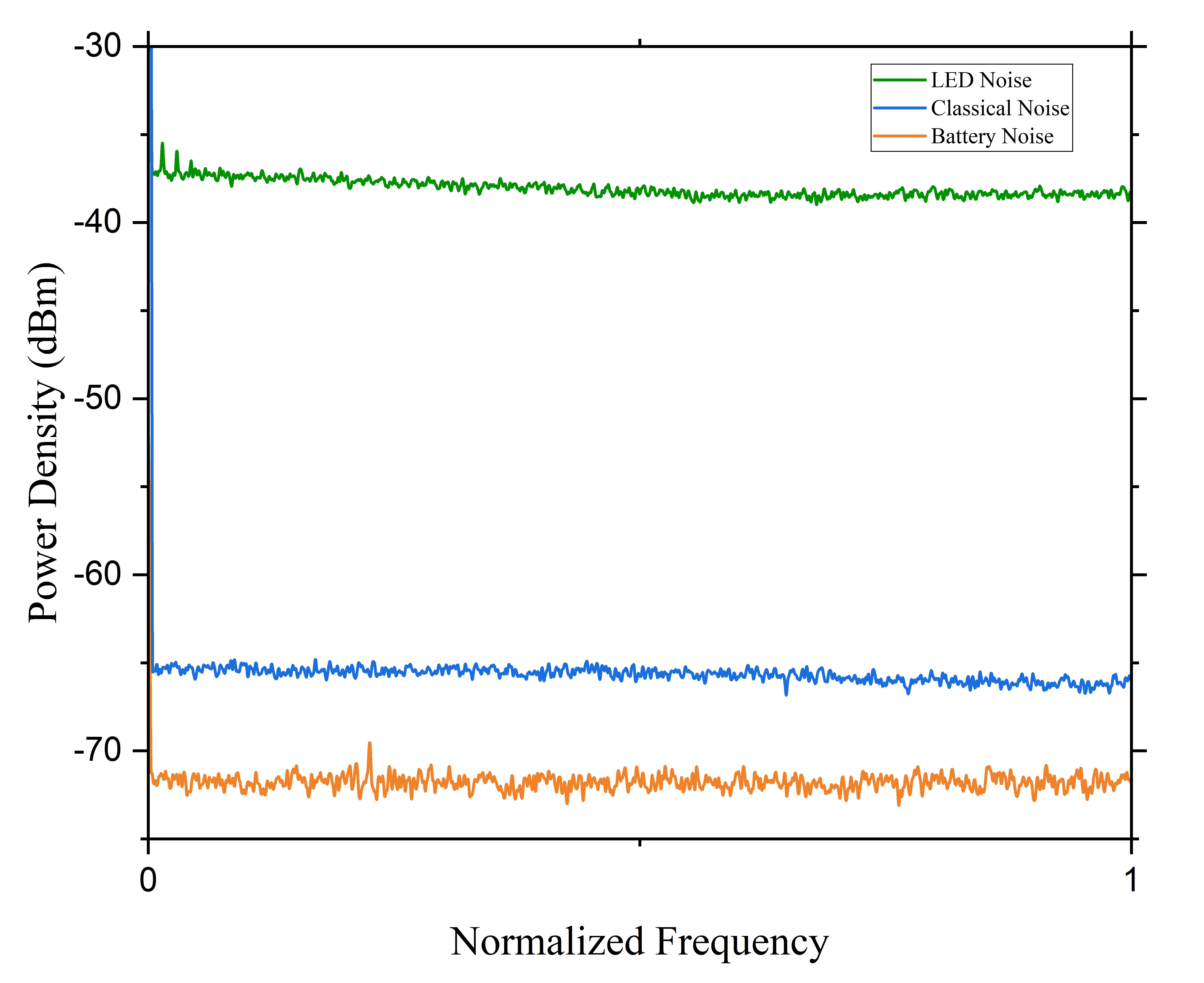}}
		\end{subfigure}
	\begin{subfigure}[]{
		\label{frequency_spec}
		\includegraphics[scale = 0.35]{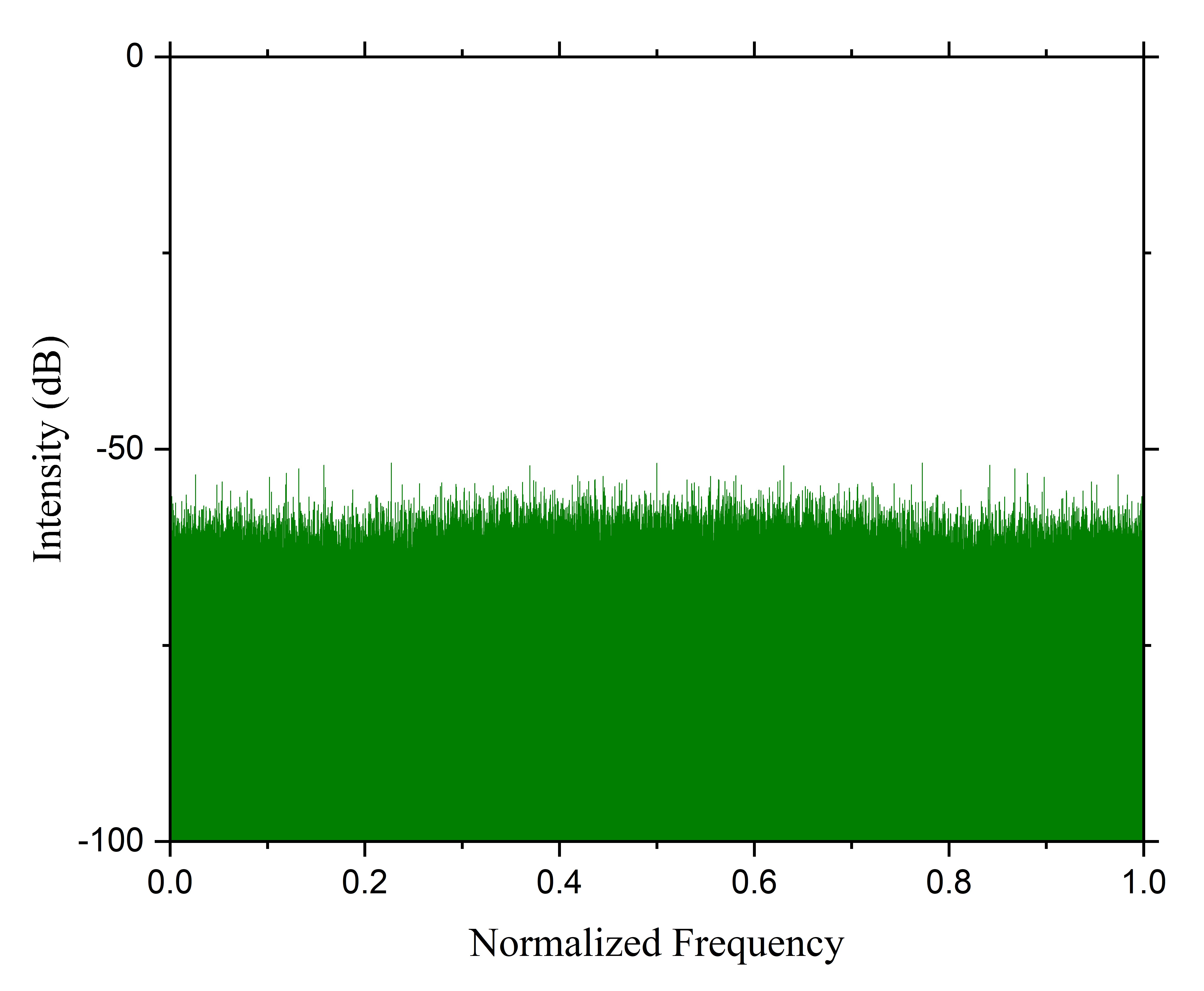}}
		\end{subfigure}

\caption{(a) Temporal waveforms of the LED and amplifier circuit's quantum signal and classic noise, where quantum noise is the prevailing factor. The green curve indicates the LED noise, and the blue curve represents the amplified classical noise, and the orange curve illustrates the amplified battery noise. (b) The histogram distribution of the signal voltage is represented by the green bars, with the dashed lines indicating a well-fitted Gaussian distribution. (c) The spectral power density in three different conditions: the LED is off (classical noise), the LED is on (LED noise), and the battery. (d) The intensity noise indicates white noise when the LED is turned on. }
\label{fig_time_wave_hist}
\end{figure*}

So far we have discussed the theoretical foundation of our QRNG device, and FIG.~\ref{fig_Fermi_recombinations} illustrates a conceptual mechanism of spontaneous emission in an LED. We will now describe the setup of the experiment. The PCB circuit of the integrated optical and electronic components and the schematic process are shown in FIG.~\ref{fig_QRNG} and ~\ref{fig_Schematic}, respectively. The detailed design block diagram of the QRNG module comprises three main parts. First, the quantum entropy source (LED). Second, the amplification of the signal received from LED by a photodetector (PD) and an amplifier (AMP). Finally, digitizing the amplified signal by an Analog-to-Digital Converter (ADC) with 12-bit resolution and 4 MSa/s sample rate and applying a post-processing procedure using a microprocessor. As demonstrated in FIG.~\ref{fig_time_wave}, the temporal waveforms of signal and noise are measured, indicating that the LED noise is dominant and the output signal oscillates irregularly and shows good randomness of intensity fluctuation. The signal distribution is given by the green histogram in FIG.~\ref{fig_hist_wave}, and the Gaussian distribution and symmetry are demonstrated by comparing them with the red fitting curve~\cite{Sanguinetti}. Also, FIG.~\ref{FFT_Power}, illustrates the spectral noise for three different situations: the power supply, the PD output when the LED is off (background noise), and the PD output when the LED is on. As can be seen, the LED noise is more dominant than the other noises. FIG.~\ref{frequency_spec} displays the intensity noise of the PD output voltage when the LED is on, indicating white noise.

To obtain high-quality random numbers, we should maximize the quantum part of the signal~\cite{Dynes}. In this way, we aim to achieve an SNR formula for controlling the quantum part using a measurable variable. In the following, we derive the SNR based on the LED current.

\begin{figure*}
\centering
\begin{subfigure}[]{
	\centering
	\label{St_dev}
	\includegraphics[scale = 0.38]{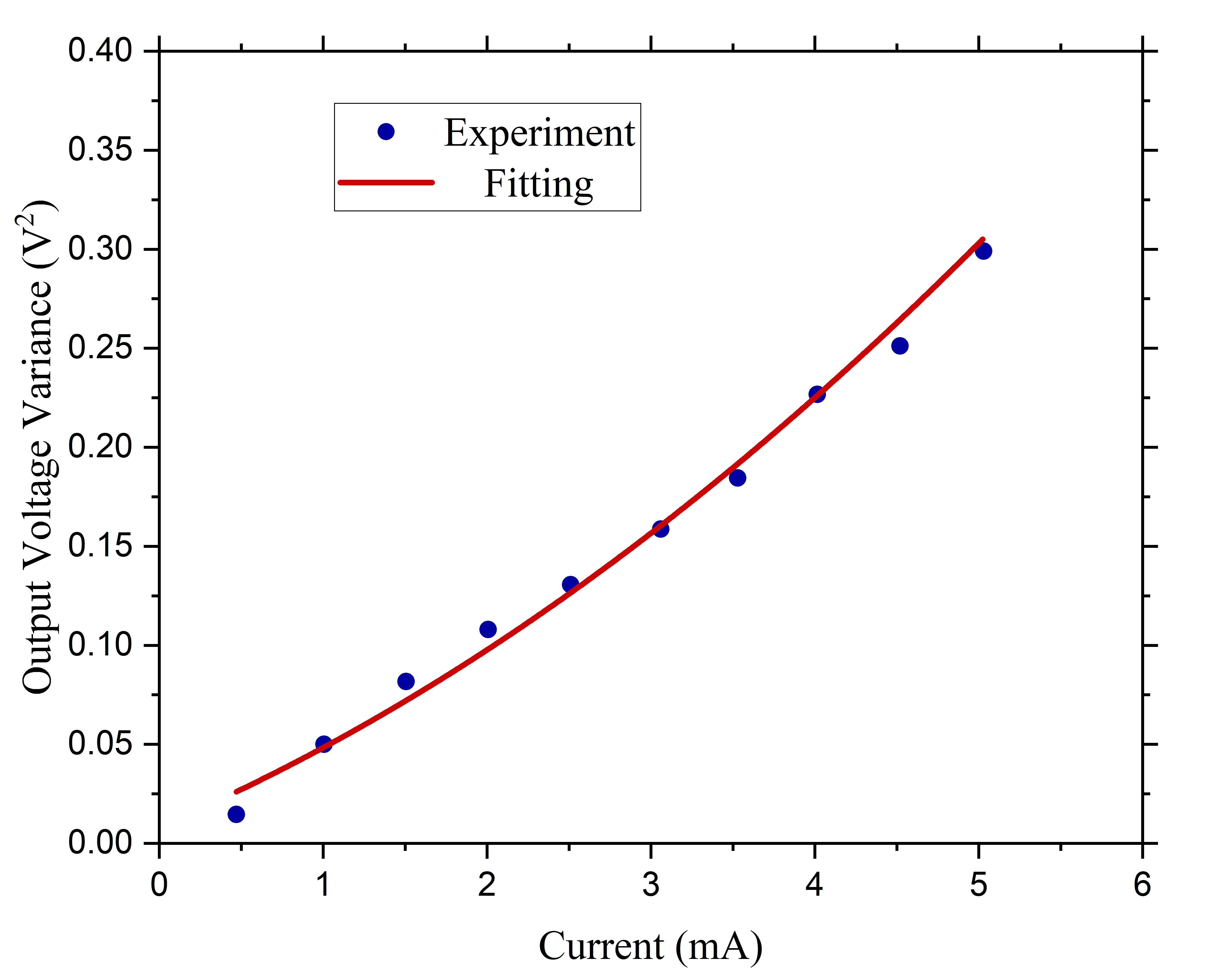}}
	\end{subfigure}
\hspace*{0.1cm}
	\begin{subfigure}[]{
		\centering
		\label{fig_SNR}
		\includegraphics[scale = 0.38]{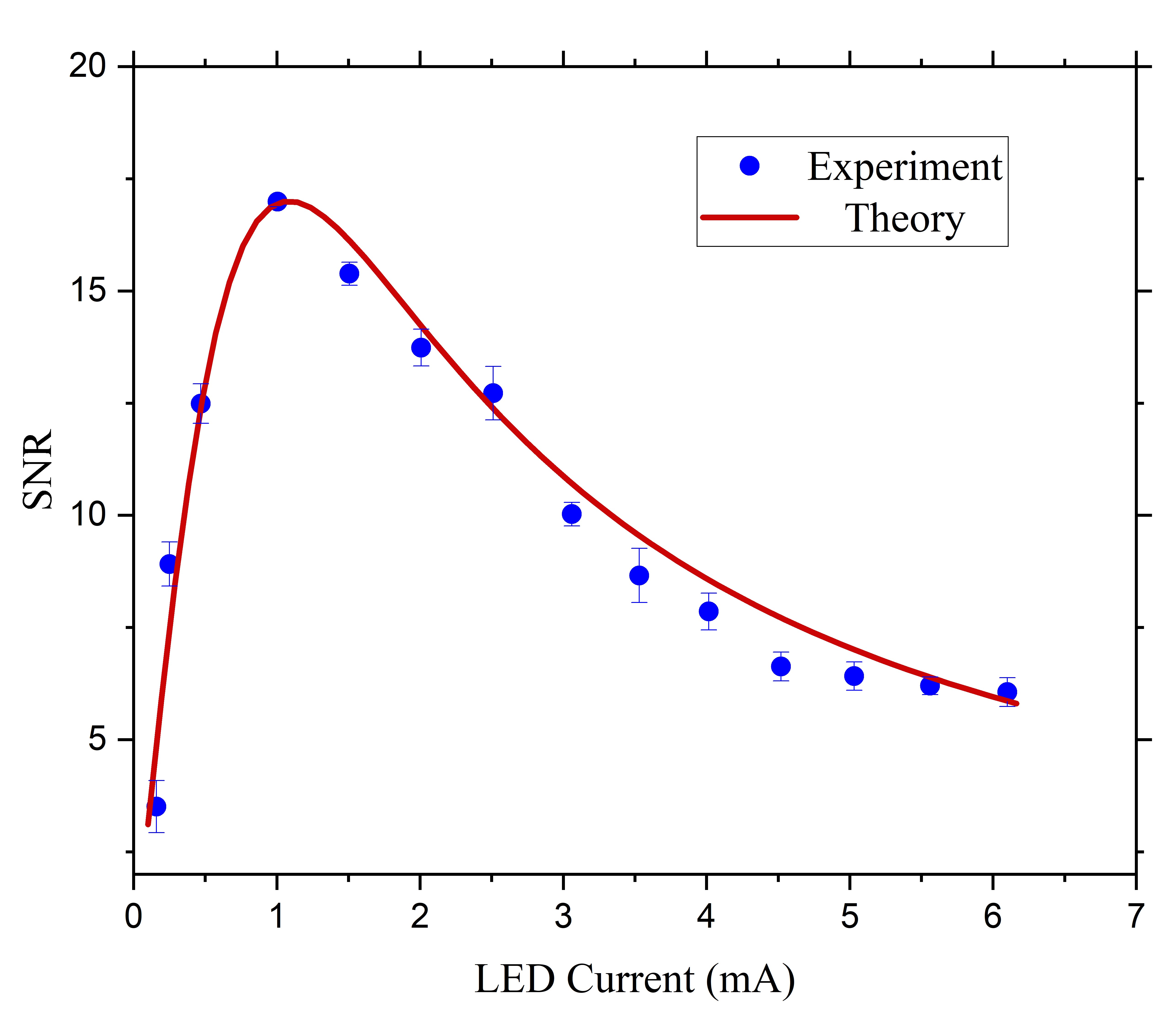}}
		\end{subfigure} 
\caption{(a) Photodetector voltage variance. The $V_{\PD}$ variance is measured by a 12-bit ADC and fitted with the polynomial function described in Eq.~(\ref{Eq_voltage_variance_current}). (b)The figure presents the ratio of quantum signal to classical noise, revealing that as the LED current decreases, the SNR increases as per theoretical predictions. When the LED turns off, background noise dominates the quantum signal, causing a rapid drop from 1 to 0 mA.} 
\end{figure*}

The electric field inside the active region is described by
\begin{align}
\label{Eq_electric_field}
E(t) = E_{0} \EXP(i(\omega t + \phi(t))
\end{align}
where $\phi$ represents the phase fluctuations of the LED executes a Brownian motion and has a Gaussian probability distribution~\cite{Henry}.

 It should be noted that quantum phase fluctuations are inversely proportional to the light intensity, whereas classical noises (e.g. flicker noise and occupation fluctuation) are independent of power~\cite{Henry,Vahala}. Thus, the total phase fluctuations can be written as follows~\cite{Henry,Xu}:
\begin{subequations}
\begin{equation}
\label{Eq_phase_variance}
\langle \phi (t)^{2} \rangle = \frac{Q}{I} + C 
\end{equation}
\begin{equation}
Q = \frac{R_{\Spon}t}{2}
\end{equation}
\end {subequations}
The contributions of quantum phase noise and classical phase noise are represented by $\frac{Q}{I}$ and C, respectively and $ \langle \bullet \rangle $ represents a statistical average. The variance of the photodetector AC voltage ($V_{PD}$) is given by
\begin{align}
\label{Eq_voltage_var}
\langle V_{PD}(t)^{2} \rangle = gQI + gCI^{2} + F
\end{align}
where $g$ and F are the gain of the photodetector and the background noise, respectively. Using Eqs~(\ref{Eq_Beer_Lambert_appx}),(\ref{Eq_voltage_var}) and converting the light intensity to current ($I_{0} = J \eta_{\ex}$), we can derive
\begin{subequations}
\begin{equation}
\label{Eq_voltage_variance_current}
\langle V_{PD}(t)^{2} \rangle = A_{QS} J + A_{\CS} J^{2} + F
\end{equation}
\begin{equation}
A_{\QS} = g Q(1-\alpha L_{\eeff})\eta_{\ex}
\end{equation}
\begin{equation}
A_{\CS}= gC(1-\alpha L_{\eeff})^{2}\eta_{\ex}^{2}
\end{equation}
\end {subequations}
where $\eta_{\ex}$ and $J$ are the external quantum efficiency and LED current, respectively.  As mentioned earlier, we need to maximize the quantum part of the signal over the classical part to generate high-quality random numbers, so we define the SNR equation~\cite{Xu}
\begin{align}
\label{Eq_SNR}
SNR = \frac{A_{\QS}J}{A_{\CS }J^{2} + F}
\end{align}
which shows the proportion of the quantum part of the signal over the classical part. As shown in FIG.~\ref{St_dev}, $A_{\QS}$, $A_{\CS}$, and $F$ were determined experimentally by measuring the variance of the photodetector voltage at different currents. The denominator represents the overall classical noise, which has two parts. Flicker noise and occupation noise from LEDs, as well as background noises (e.g. shot noise, flicker noise, and thermal noise). Aside from that, background noise is independent of LED current and is determined by the amplifier bias condition. In addition, we have used a BJT amplifier to reduce the Flicker noise of the amplifier circuit. The results demonstrate that at low and high current ranges, either background noise or classical noise will dominate the quantum signal, while at medium currents (1 mA to 2 mA for our setup), the quantum signal will dominate, as shown in FIG.~\ref{fig_SNR} and predicted by Eq.~(\ref{Eq_SNR}). Therefore, we determined the optimal current range where the quantum signal is the strongest and the classical noise is minimal.

\begin{figure*}
\centering
\begin{subfigure}[]{
\centering
\includegraphics[scale=0.27]{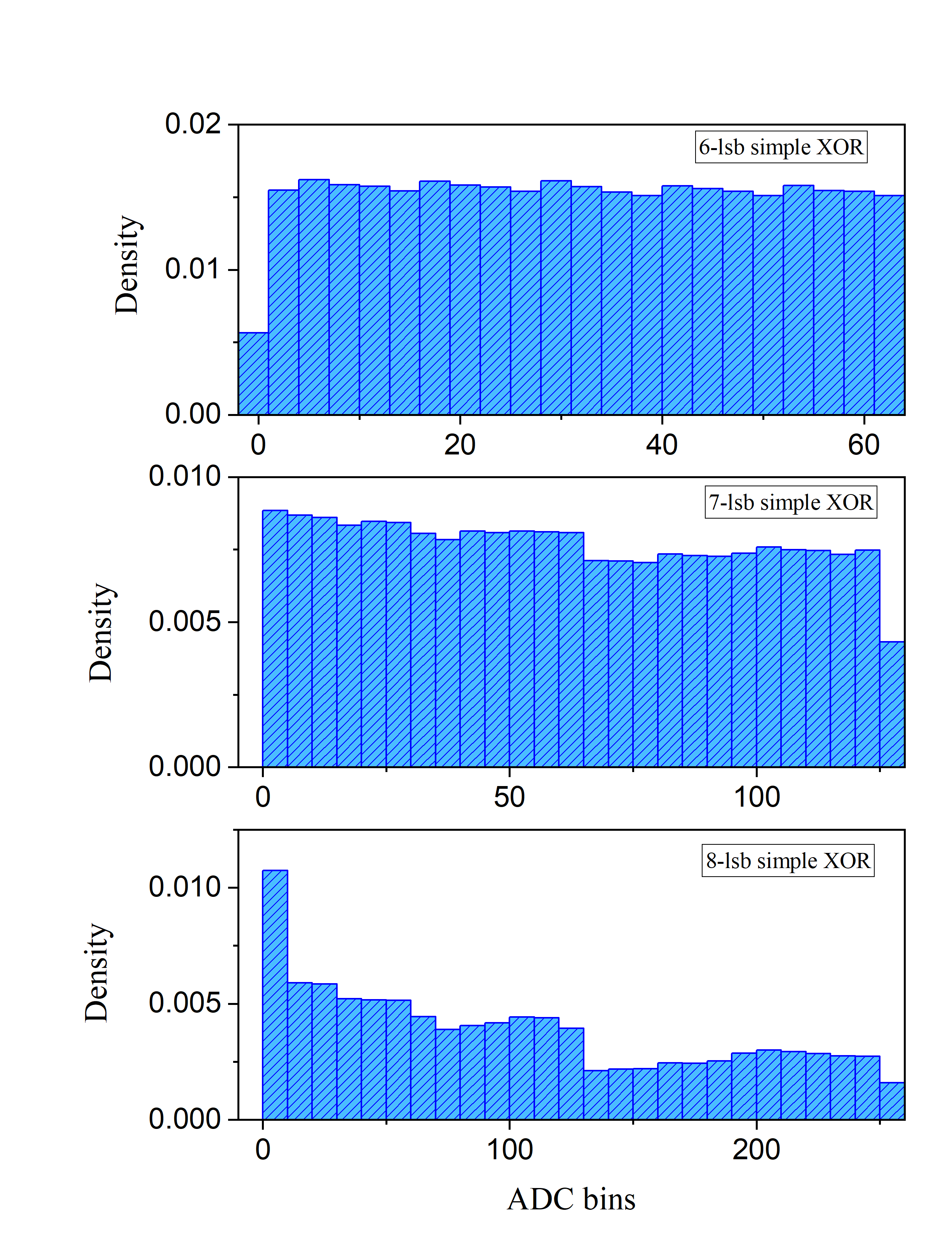}}
\end{subfigure}
\begin{subfigure}[]{
\centering
\includegraphics[scale=0.27]{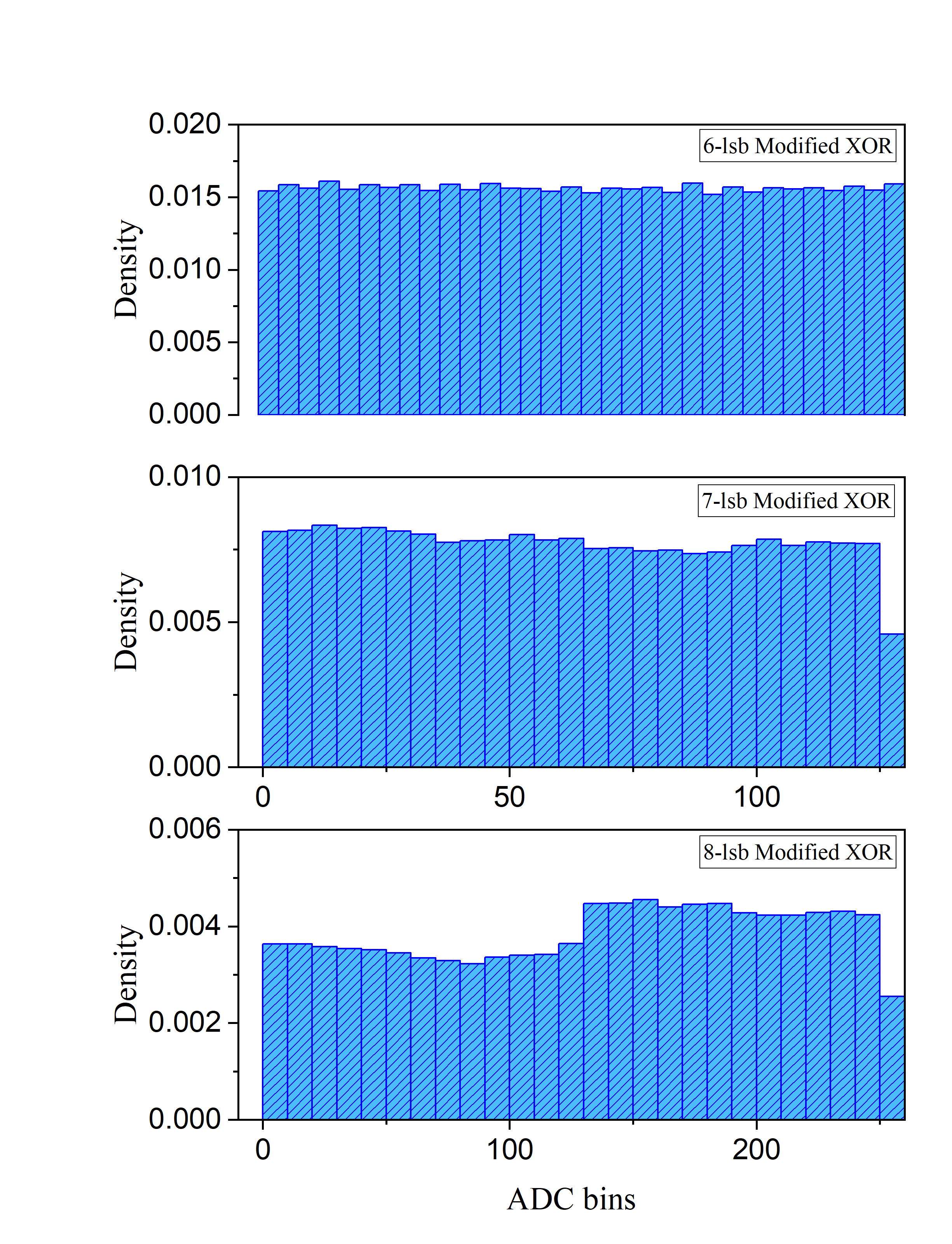}}
\end{subfigure} 
\begin{subfigure}[]{
\centering
\includegraphics[scale=0.27]{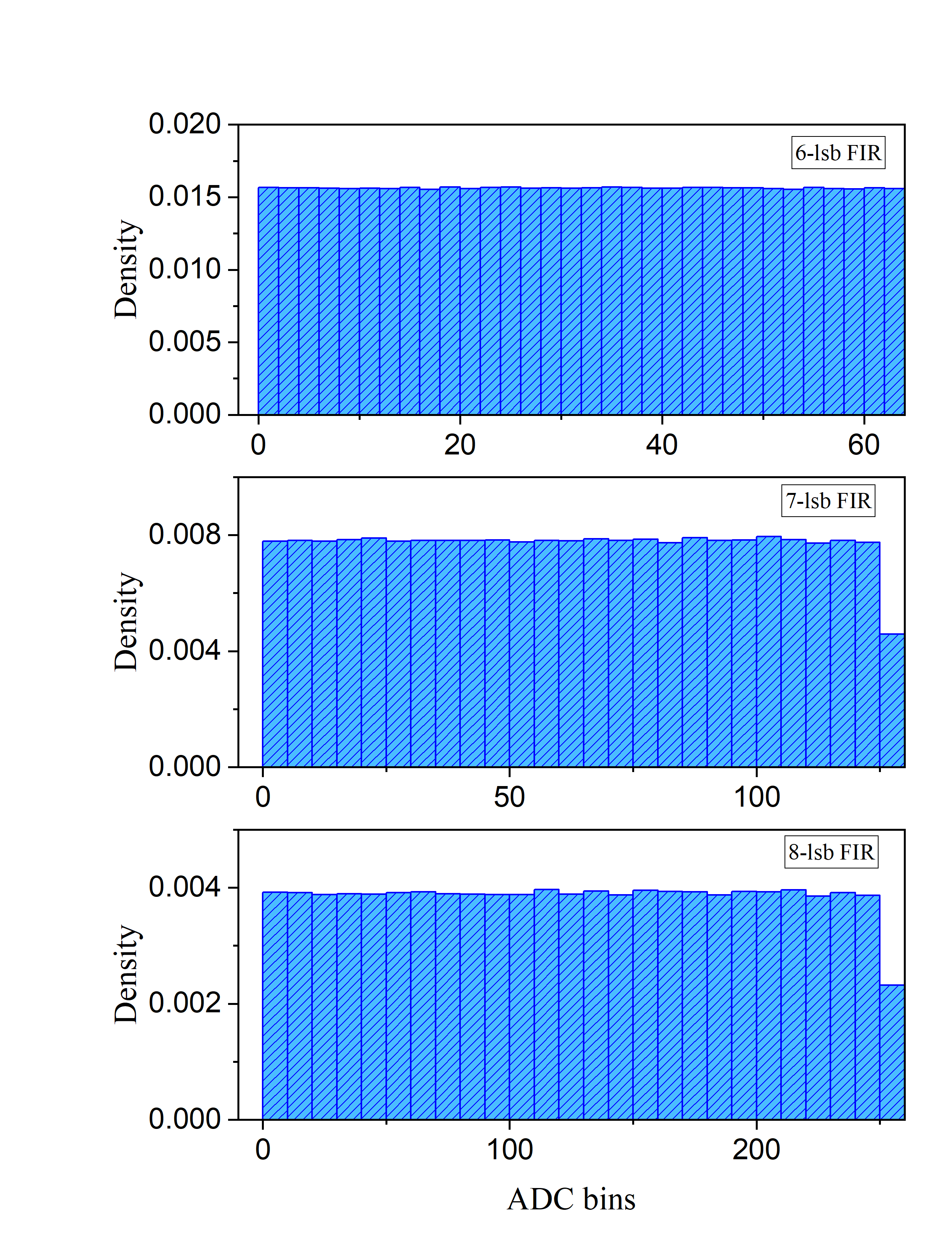}}
\end{subfigure}
\caption{The probability distribution of three post-processing approaches with different m-lsb values (m=6, m=7, m=8): (a) Simple XOR, (b) Modified XOR, and (c) FIR. The FIR method produces a perfectly uniform distribution, while the other two approaches exhibit some non-uniformity.}
\label{fig_mLSBs}
\end{figure*}

\begin{figure} 
\centering
\includegraphics[scale=0.36]{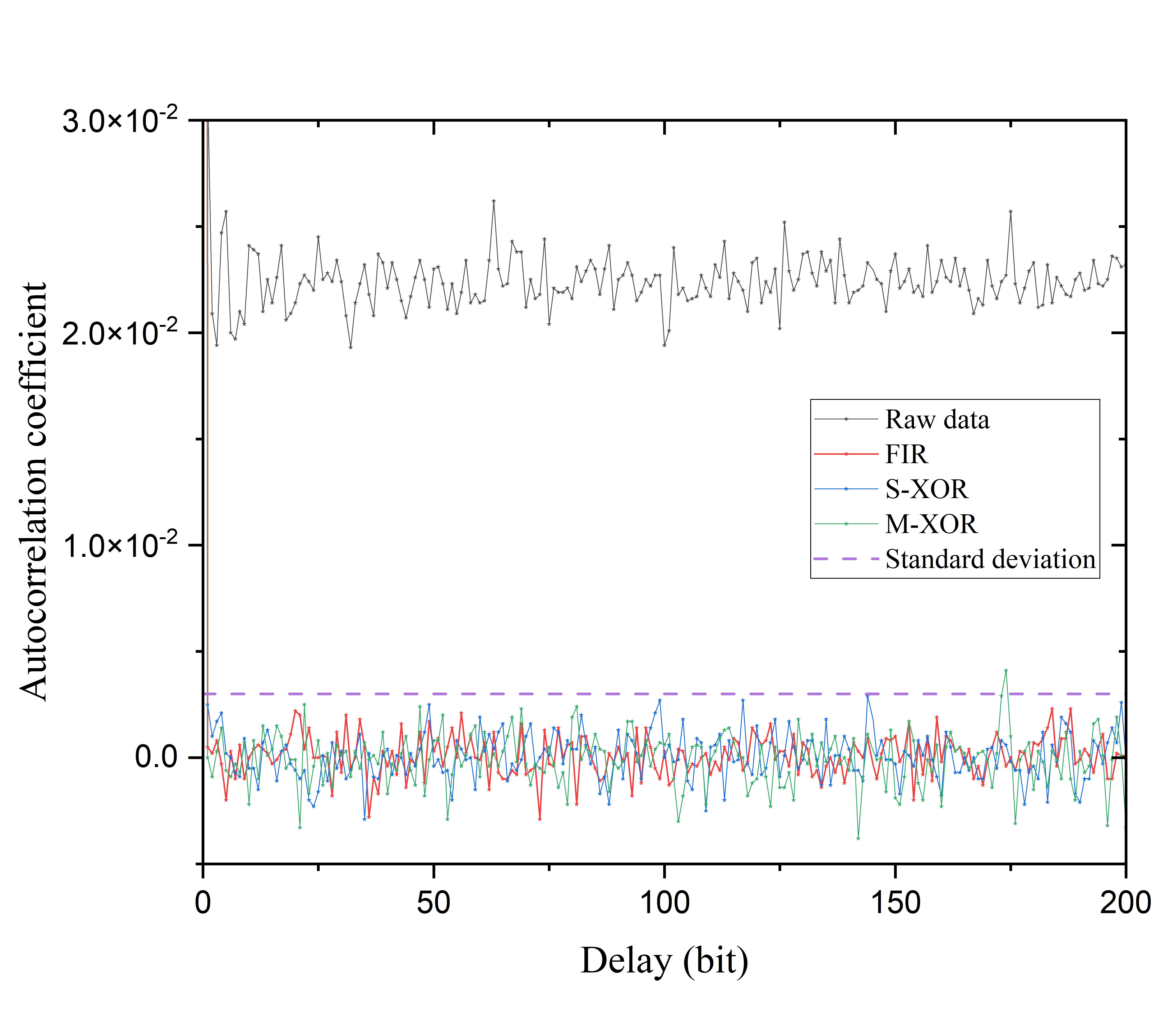}
\caption{The autocorrelation coefficient for four different datasets: raw data, S-XOR, M-XOR, and FIR. The standard deviation is denoted by the dashed line.}
\label{fig_Autocorrelation}
\end{figure}

\begin{figure*}
\centering
\begin{subfigure}[]{
	\centering
	\label{fig_min_entropy_time}
	\includegraphics[scale = 0.35]{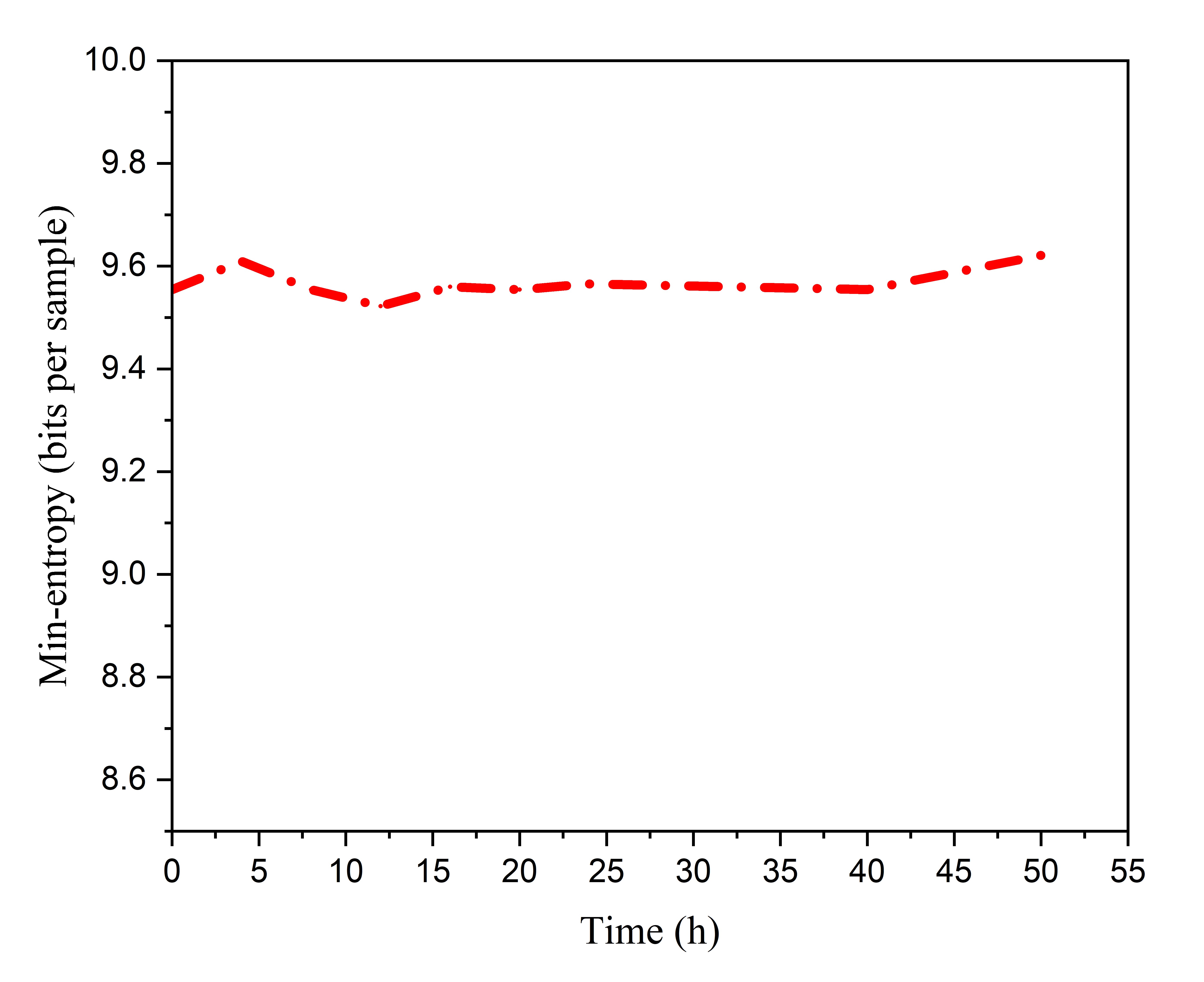}}
	\end{subfigure}
\hspace*{0.1cm}
	\begin{subfigure}[]{
		\centering
		\label{fig_min_entropy_temp}
		\includegraphics[scale = 0.35]{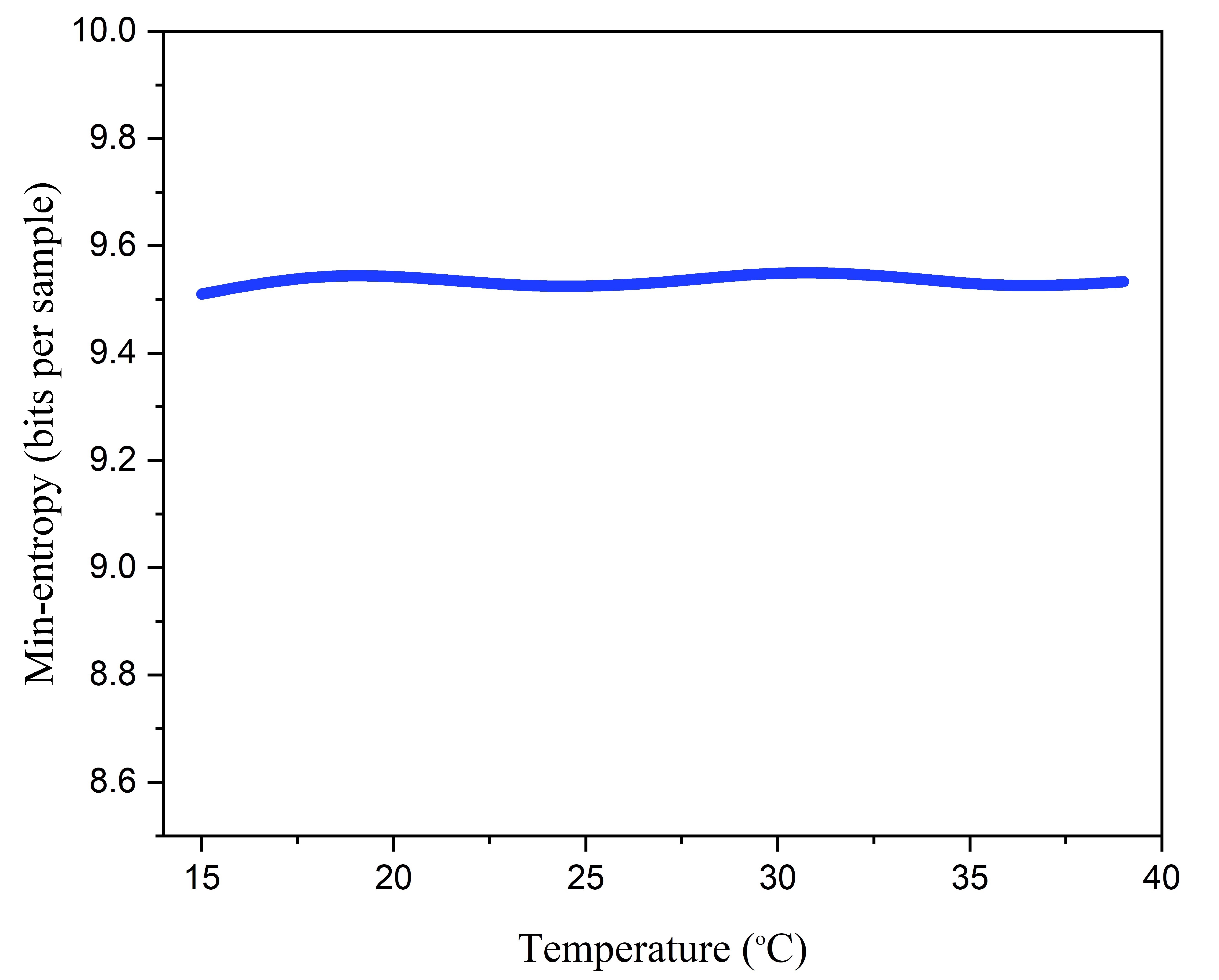}}
		\end{subfigure} 
\caption{The figure depicts how the min-entropy changes over time and at different temperatures, indicating the device's dependable and consistent production of randomness over time (a) and different temperatures (b).} 
\end{figure*}

Typically, QRNGs use randomness extractors such as Toeplitz hashing to generate information-theoretically provable random numbers. However, randomness extractions are too computationally intensive, especially for real-time embedded devices based on microprocessors. This will significantly decrease generation rate. We used simple post-processing suitable for microprocessors to avoid hardware complicity and low-speed generation rate. Aside from that, the purpose of standard RNGs is to generate a random uniform string, in which the raw numbers are processed to obtain a good-quality output with a uniform distribution. To achieve this, we explored three methods suitable for embedded real-time devices to convert the Gaussian distribution of the QRNG into a uniform distribution. 
The first approach involved a simple XOR (S-XOR) processing technique, where a single number from the ADC was XORed with the $k_{\thh}$ number following it. The second approach was the modified XOR (M-XOR) method, where the output of the ADC was XORed with a shift-rotated version of itself and then XORed with the $k_{\thh}$ number following it. The last approach is the finite impulse response (FIR) method~\cite{Ifeachor}. The FIR method involves taking a weighted sum of past input samples, as described by Eq.~(\ref{eq_FIR}), to analyze the input signal. This can help to reduce any unwanted noise or interference in the signal, leading to a higher-quality output. 
Studies have shown that the FIR method can also maintain the min-entropy of the generated random numbers. As a result, this method creates a more unpredictable result by scrambling the raw input data~\cite{Marangon}.
This technique consists of transforming a raw integer sample $x(n)$ into an unbiased one $y(n)$, by means of the relation: 
\begin{align}
\label{eq_FIR}
y(n) = \sum_{i = 0}^{M} b_{i}x(n - i) 
\end{align}
where $b_{i} = \frac{M!}{i!(M-i)!}$ and M is the number of raw samples. After every technique, we have taken m-Least Significant Bit (m-LSB).

To establish a robust and reliable device we compare and evaluate three different post-processing procedures for real-time embedded systems that were introduced in the previous section. These methods were compared according to their processing time, their ability to maintain minimum entropy, and their ability to produce uniformly distributed random numbers that passed the National Institute of Standards and Technology (NIST) randomness test~\cite{NIST1,NIST}. Min-entropy is defined as 
\begin{align}
\label{eq_min_entropy}
H_{\infty}(x) = -\log_2 (\max(\Prr[x=X]))
\end{align}
which indicates how random a distribution $X$ is on $\{0,1\}^n$. The findings of this study are presented in Figs. \ref{fig_mLSBs} and \ref{fig_Autocorrelation}, which illustrate the m-lsb values obtained through the application of three distinct post-processing methods, as well as the autocorrelation between the resultant data. As it has been shown in FIG.~\ref{fig_Autocorrelation}, the autocorrelation of post-processed data is below the standard deviation means the random data are not correlated~\cite{Xu}. The FIR method produced a uniform probability distribution at 10-lsb and passed the NIST randomness test, while the S-XOR and M-XOR methods produced uniformity at 5-lsb and 6-lsb, respectively. Table~\ref{Tab_NIST} presents the results of the NIST randomness test for each approach. Our findings indicate that the S-XOR approach did not pass some of the NIST tests, as presented in Table~\ref{Tab_NIST}. Therefore, caution is necessary when relying solely on the uniform distribution achieved by the S-XOR approach.

\begin{table}

\caption{\small Performance comparison of NIST results for S-XOR, M-XOR, and FIR methods.} 
\centering
\setlength{\tabcolsep}{5pt} 
\begin{tabular}{c c c c c c c c c c c c c c c c c}
\hline \hline
{\bf Statistical Tests} & {\bf FIR} & {\bf M-XOR} & {\bf S-XOR}\ \\ 
\hline \hline
Frequency & Pass & Pass & Fail\\
Block Frequency & Pass & Pass & Fail\\
Runs & Pass & Pass & Fail \\
Longest Run & Pass & Pass & Fail\\
Rank & Pass & Pass & Pass \\ 
FFT & Pass & Pass & Pass\\ 
Non-Overlapping Template & Pass & Pass & Pass\\ 
Overlapping Template & Pass & Pass & Pass \\ 
Universal & Pass & Pass & Pass \\ 
Linear Complexity & Pass & Pass & Pass \\ 
Serial & Pass & Pass & Pass \\
Approximate Entropy & Pass & Pass & Pass \\
Cumulative Sums & Pass & Pass & Fail \\
Random Excursions & Pass & Pass & Fail \\
Random Excursions Variant& Pass & Pass & Fail \\
\hline \hline
\end{tabular}
\label{Tab_NIST}
\end{table}

The results assert that the FIR method is the most suitable approach for post-processing in real-time embedded systems. This method has been found to maintain the min-entropy of the random data, thereby improving the reliability and robustness of the device. Following a comparative analysis of three alternative methods, we have selected the FIR method for use in our device. To further verify the performance of our device, we conducted two tests to evaluate the performance of our device, including temperature tests to determine the impact of temperature on device performance and time duration tests to check the stability of device functionality over time. These tests have provided valuable insights into the effectiveness and reliability of the FIR method for post-processing in real-time embedded systems. 
FIGs.~\ref{fig_min_entropy_time} demonstrates the variation of the min-entropy of random data over time. The results show that the min-entropy remains almost constant, indicating that the fabricated device is robust and reliable over time. In addition, a temperature test was conducted to evaluate the device's durability and ability to withstand temperature variations. The results, illustrated in FIG.~\ref{fig_min_entropy_temp}, demonstrate that the device performs well and maintains its reliability and robustness even when subjected to temperature fluctuations.

Finally, our QRNG device now reaches a generation rate of 1 Mb/s and has the potential to attain higher rates by integrating additional LEDs in a parallel configuration.

\mycomment{
\begin{figure}
\centering
\includegraphics[scale=0.085]{qrngPIC.jpg}
\caption{Schematic of the built QRNG. The enclosure size is 16 × 12× 6 $\ccm^{3}$.}
\label{fig_QRNG}
\end{figure}
}

In summary, we have fabricated a durable, low-power, cost-effective QRNG uses LED spontaneous emission. Results show that quantum signals dominate when the current is medium-to-low. Our research attempts to establish a link between absorption and LED temporal fluctuations as another source of quantum noise in an LED. Moreover, we evaluated various post-processing methods and discovered that the FIR approach is the most reliable and yields the highest min-entropy outcomes. Our device maintains a stable min-entropy over time and under variable temperatures, operates at a real-time rate of 1 Mb/s, and has passed all the NIST tests. These results provide a promising route for building efficient and practical QRNGs with potentially high bit rates.

\mycomment{
\begin{acknowledgments}
.....................
\end{acknowledgments}
}

\end{document}